\documentclass[aps,onecolumn,eqsecnum,amsmath,nofootinbib,preprintnumbers]{revtex4}%

\usepackage[dvips]{color,graphicx}
\usepackage{amsfonts,amssymb,theorem,mathrsfs,times}
\usepackage{CJK}   
\usepackage{color} 
\begin{document}
\begin{CJK*}{GBK}{com}  

\preprint{\hfill {\small {ICTS-USTC-13-01}}}

\title{ Maximum Entropy Principle for Self-gravitating Perfect Fluid in Lovelock Gravity}

 \author{ Li-Ming Cao\footnote{e-mail address:
caolm@ustc.edu.cn}, Jianfei Xu\footnote{e-mail address:
jfxu06@mail.ustc.edu.cn} and Zhe Zeng\footnote{e-mail address:
zengzhe@mail.ustc.edu.cn}\\}


\address{
\\Interdisciplinary Center for Theoretical Study \\
University of Science and Technology of China\\
Hefei, Anhui 230026, China\\}

\vspace*{2.cm}
\begin{abstract}
We consider a static self-gravitating system consisting of perfect fluid with isometries of an $(n-2)$-dimensional maximally symmetric space in  Lovelock gravity theory. A straightforward analysis of the time-time component of the equations of motion suggests a generalized mass function. Tolman-Oppenheimer-Volkoff like equation is obtained by using this mass function and gravitational equations. We investigate the maximum entropy principle in Lovelock gravity, and find that this Tolman-Oppenheimer-Volkoff equation can also be deduced from the so called ``maximum entropy principle" which is originally customized for Einstein gravity theory. This investigation manifests a deep connection between gravity and thermodynamics in this generalized gravity theory.
\end{abstract}
\maketitle

\section{Introduction}

The relation among gravity, thermodynamics and quantum theory has been studied for about half a century.
Starting from Einstein equations and appropriate definitions of mass, angular-momentum and charges, people have found
the four laws of black hole mechanics~\cite{bardeen1973}. Once the quantum field on the spacetime of a black hole
is considered, it is found that black hole behaves like a black body with a temperature which is proportional to the surface gravity
at the horizon of the black hole~\cite{hawking}. Then, the black hole mechanics is naturally interpreted as black hole thermodynamics with
an entropy which is one quarter of the area of the horizon~\cite{Bekenstein}.
Since these seminal works, the thermodynamics
of black holes and other possible spacetimes in various gravity theories have been widely discussed, and now people believe
that the black hole is a real thermodynamical system.

Actually, for a given gravity theory, assuming some spacetime which contains horizon(s) and has a good asymptotical behavior,
then, based on the related gravitational equations, we could study the corresponding conserved charges, the geometry of the horizon and
some possible quantum effects, and then develop the mechanics or thermodynamics of this spacetime. For example, the
thermodynamics of stationary black holes in general differmorphism invariant gravity theory has been established~\cite{Wald:1999vt}(references therein).
This is a traditional way to study the thermodynamics of the spacetime since the beginning of 1970s. On the other hand, some people believe that the gravitational equations (or a part of the gravitational equations) can also be derived from some fundamental relations in thermodynamics.
Once the logic is turned  around, some problems arise: For instance, in the traditional discussion of thermodynamics of spacetime,
the globally defined conserved charges (such as ADM mass) suggests that it is quite difficult to extract useful local information from
the law of thermodynamics. So, to solve such problem, one has to introduce some locally or quasilocally defined thermodynamic quantities and consider
the local or quasilocal version of the thermodynamic relations.  The  gravitational equations (or some part of these equations), such as Einstein equations, are really
possible to be derived from these local or quasilocal thermodynamic quantities and associated thermodynamic relations. In fact, based on the definition of local
Rindler horizons, Jacobson shows that the Einstein equations can be derived from the Clausius relation~\cite{Jac,Jac1}.
In the case with enough isometries, based on a particular  quasilocal mass, i.e., the so called Misner-Sharp energy~\cite{Misner:1964je,Hayward:1994bu} (a typical quasilocal energy), one can also derive a part of Einstein equation from the first law or Clausius relation of thermodynamics~\cite{c1,c2,c3,gong}.

For static self-gravitating fluid systems, maximum entropy principle can be used to deduce a part of the Einstein equations,
i.e, Tolman-Oppenheimer-Volkoff (TOV) equation. This principle is first applied by Cocke~\cite{cocke} to the self-gravitating
perfect fluid sphere and improved by Sorkin, Wald and Zhang (SWZ)
~\cite{wald}.
In the discussion of SWZ, only the time-time component of the Einstein equations is used.
However, SWZ's discussion was restricted to radiation. Recently, Gao has generalized SWZ's discussion to an arbitrary perfect fluid, and has successfully got the usual TOV equation in Einstein theory by the principle of maximum entropy~\cite{gaosijie}.
In these discussions, the existence of a ``mass function" $m(r)$ is quite important. Actually, it
is just the Misner-Sharp energy applying to this static system~\cite{Hayward:1994bu}. This mass function is very special,
and it only exists in some special gravity theories, such as Einstein gravity and Lovelock gravity~\cite{Love}.

The maximum entropy principle clearly and profoundly reveals the deep connection between thermodynamics and gravity.
However, at present, only Einstein gravity theory is considered. Our question is: Whether this principle is valid or not in more general gravity theories?
In this paper, we will consider a static self-gravitating
system consisting of a perfect fluid with isometries of an $(n-2)$-dimensional maximally symmetric space in  Lovelock
gravity theory. Firstly, generalized TOV equation in this theory is presented, then, we show that the maximum entropy principle can also be
applied to this self-gravitating perfect fluid system: Correct TOV equation can be obtained by a similar deduction as in~\cite{gaosijie}.
Our results show that the implicit relationship between gravity and
thermodynamics in Einstein gravity still exists in Lovelock theory, which manifests the possibility of
turning the logic around to use thermodynamics to describe the more general gravitational phenomena.

This paper is organized as follows: In Sec.2, we give a brief review of  the Lovelock gravity.
In Sec.3, starting from the equations of motion of the Lovelock gravity, we introduce a generalized mass function for the system with isometries of an $(n-2)$-dimensional maximally symmetric space. Based on this mass function, generalzied TOV equation is obtained.
In Sec.4, we apply the maximum entropy principle to this system to derive the TOV equation.
Section 5 is devoted  to the conclusion and discussion.

\section{Lovelock gravity}
\label{Lovelock}

Lovelock gravity~\cite{Love} is a natural generalization of general relativity in higher dimensions
in the sense that the equations of motion of Lovelock gravity do
not contain more than second order derivatives with respect to
metric, as the case of general relativity.
Actually, with the development of string theory and supergravity over the past years,
Lovelock gravity has drawn a lot of attention. It is well known that Einstein general relativity naturally arises from
string theories. As corrections from massive states of string
theories and from loop expansions in string theories, the higher order curvature terms in lovelock theory  appear in the low-energy effective
action of string theories~\cite{Zwie, Lowenergylimit, Lowenergylimit1,Lowenergylimit2,Lowenergylimit3}. So it is important to investigate the
effects of these higher order terms. For example, these effects have been manifested in black hole solutions~\cite{Boulware:1985wk,Cai:2001dz}.

The Lagrangian density of Lovelock gravity consists of the dimensionally
extended Euler densities which are given by
\begin{equation}
L=\sum^{[n/2]}_{i=0}\alpha_{i}L_{i} \,,
\end{equation}
where $\alpha_{i}$s are constants and $L_{i}$s are the Euler densities of
$2i$-dimensional manifolds defined by
\begin{equation}
L_{i}=\frac{1}{2^i}\delta^{a_{1}b_{1}...a_{i}b_{i}}_{c_{1}d_{1}...c_{i}d_{i}}R^{c_{1}d_{1}}_{\phantom{1}\phantom{1}
a_{1}b_{1}}...R^{c_{i}d_{i}}_{\phantom{1}\phantom{1} a_{i}b_{i}} \, .
\end{equation}
The symbol $\delta$  denotes a totally anti-symmetric product of Kronecker deltas  defined as
\begin{equation}
\delta^{a_{1}...a_{p}}_{b_{1}...b_{p}}=p!\delta^{a_{1}}_{
[b_{p}}...\delta^{a_{p}}_{b_{p}]} \,.
\end{equation}
The action for Lovelock gravity is given by
\begin{equation}
S=\frac{1}{2\kappa^{2}_{n}}\int
d^{n}x\sqrt{-g}\sum^{[n/2]}_{i=0}\alpha_{i}L_{i}+S_{\mathrm{matter}} \,.
\end{equation}
Following from the variation of this action, the gravitational equation is written in a compact form
\begin{equation}
\label{GT}
\mathcal{G}_{ab}=\sum_{i}^{[n/2]}\alpha_{i}G^{(i)}_{ab}=\kappa^{2}_{n}T_{ab} \,,
\end{equation}
where
\begin{equation}
G^{(i)f}_{~~e}=-\frac{1}{2^{i+1}}\delta^{fa_{1}b_{1}...a_{i}b_{i}}_{ec_{1}d_{1}...c_{i}d_{i}}R^{c_{1}d_{1}}_{\phantom{1}\phantom{1}
a_{1}b_{1}}...R^{c_{i}d_{i}}_{\phantom{1}\phantom{1} a_{i}b_{i}}
\end{equation}
is deduced from the variation of $L_{i}$, and $T_{ab}$ is
the energy-momentum tensor for matter fields obtained from the matter action
$S_{\mathrm{matter}}$. In the case where only $\alpha_0$ and $\alpha_1$ are nonvanishing, we get usual Einstein gravity with the cosmological constant. If $\alpha_2$
is also nonvanishing, we get the Gauss-Bonnet gravity theory.

\section{TOV Equations in Lovelock Gravity Theory}
\subsection{static perfect fluid in Lovelock gravity}
Let us consider a perfect fluid system with symmetries corresponding to the isometries of an $(n-2)$-dimensional maximally symmetric space in Lovelock gravity. Additionally, we assume the system is static (an additional isometry). The spacetime of the system automatically has such isometries. By choosing coordinates $\{t,r,z^i\}$ with $i=,1,\cdots, n-2$, the metric of the spacetime is given by
\begin{equation}\label{1}
ds^{2}=-e^{2\Phi(r)}dt^{2}+e^{2\Psi(r)}dr^{2}+r^{2}\gamma_{ij}dz^{i}dz^{j} \, ,
\end{equation}
where $\gamma_{ij}dz^idz^j$ is the metric of an $(n-2)$-dimensional maximally symmetric manifold. Assuming the  energy-momentum tensor of the perfect fluid is given by
$$T_{ab}=\rho u_{a}u_{b}+p(g_{ab}+u_{a}u_{b})\, ,$$ where $u^a$ is a timelike vector field which is the tangent of the world line of the fluid element, then, we can further choose the time coordinate $t$ so that 
$u_{a}=-e^{\Phi}(dt)_{a}$. Actually, $u^a$ is just the vector field for the static observer (which is also the comoving observer of the fluid). The energy density $\rho$ is just measured by this observer.

It is easy to find that the nontrivial components of the Riemann tensor of such spacetime is given by
\begin{eqnarray}
&&R^{tr}_{~~tr}=-2e^{-2\Psi}(\Phi'^2-\Phi'\Psi'+\Phi'')\, ,\qquad R^{ti}_{~~tj}=\frac{e^{-2\Psi}}{r}\Phi'\delta^i_{~j}\, ,\nonumber\\
&&R^{ij}_{~~kl}=\frac{k-e^{-2\Psi}}{r^2}\delta^i_{~k}\delta^{j}_{~l}\, ,\qquad\qquad R^{ri}_{~~rj}=-\frac{e^{-2\Psi}}{r}\Psi'\delta^i_{~j}\, .
\end{eqnarray}
where $k=0,\pm 1$ corresponds to the sectional curvature of the maximally symmetric manifold.
After substituting these results into Eq.(\ref{GT}), the equations of motion for this system are given by:
\begin{equation}
\label{tt}
\kappa_n^2 \rho= \frac{1}{r^{n-2}}\frac{d}{dr}\Bigg{\{}\sum_{i=0}^{[n/2]}\frac{\alpha_i(n-2)!}{2(n-2i-1)!}r^{n-1-2i}\left(  k- e^{-2\Psi} \right)^{i}\Bigg{\}}\, ,
\end{equation}
which comes from $\mathcal{G}_{~t}^{t}=\kappa_n^2 T_{~t}^{t}$, and 
\begin{eqnarray}
\label{rr}
\kappa_n^2 p = \sum_{i=0}^{[n/2]}\frac{i\alpha_i(n-2)!}{(n-2i-1)!} \frac{e^{-2\Psi}\Phi' }{r}\left(  \frac{k-e^{-2\Psi} }{r^2} \right)^{i-1} \nonumber \\
- \sum_{i=0}^{[n/2]}\frac{\alpha_i(n-2)!}{2(n-2i-2)!}\left(  \frac{k- e^{-2\Psi}}{r^2} \right)^{i} \, ,
\end{eqnarray}
which is given by $\mathcal{G}_{~r}^{r}=\kappa_n^2 T_{~r}^{r}$. Obviously, equation $\mathcal{G}_{~t}^{r}=\kappa_n^2 T_{~t}^{r}$ is trivially satisfied. In Einstein gravity theory without cosmological constant, Eq.(\ref{tt}) suggests that we can define a mass function $m(r)$ as
\begin{equation}
m(r):=\omega_k\int_{0}^r \rho~ x^{n-2}dx\, ,
\end{equation}
such that the left of Eq.(\ref{tt}) can be explained as the change of this mass along the radial direction, i.e., $m'=\rho \omega_k r^{n-2}$. Here, $``'"$ denotes the
derivative respect to $r$, and $\omega_k$ is the volume of the maximally symmetric manifold with the sectional curvature $k$, i.e.,
$$\omega_k:=\int d^{n-2}z \sqrt{\gamma}\, .$$
Actually, Eq.(\ref{tt}) suggests that this mass function is nothing but the so called Misner-Sharp energy~\cite{Misner:1964je, Hayward:1994bu} inside the sphere with radius $r$:
\begin{equation}
\label{mass}
m(r)= \frac{(n-2)\omega_k}{2\kappa_n^2}~r^{n-3}\left(  k- e^{-2\Psi} \right)\, .
\end{equation}
Of course, here we are considering a static fluid in  $n$-dimension, so this is generalized version of the Misner-Sharp energy in this higher dimensional Einstein gravity theory (for instance, see~\cite{c3} and references therein).

In general Lovelock gravity theory, we can also define a mass function $m(r)$ for the perfect fluid, which is given by
\begin{equation}
\label{masslovelock}
m(r)=\frac{\omega_k}{2\kappa_n^2}\sum_{i=0}^{[n/2]}\frac{\alpha_i(n-2)!}{(n-1-2i)!}r^{n-1-2i}\left(  k- e^{-2\Psi} \right)^{i}\, .
\end{equation}
So Eq.(\ref{tt}) just means
\begin{equation}
\label{mrho}
\rho~ \omega_kr^{n-2} =m'\, .
\end{equation}
This is the same as the case in Einstein gravity theory. It should be emphasized here: To get Eq.(\ref{mrho}), only the time-time component of the gravitational equations is required.

More general definition of the Misner-Sharp energy in Lovelock gravity theory can be found in references~\cite{msgb} and~\cite{msll}. An explicit form of this
energy in Vaidya spacetime has been studied in~\cite{Cai:2008mh}. However, the definition of the mass function (\ref{masslovelock}) is enough to discuss the problem in this paper.

To get the TOV equation,  let us consider Bianchi identity (or loosely speaking, energy-momentum conservation equation). It is easy to find
\begin{equation}
\Big(\frac{\partial}{\partial
r}\Big)^{b}\nabla^{a}T_{ab}=(\rho+p)\Big(\frac{\partial}{\partial
r}\Big)^{b}u^{a}\nabla_{a}u_{b}+\Big(\frac{\partial}{\partial
r}\Big)^{b}\nabla_{b}p=0 \, .
\end{equation}
For the vector $u^a$ of the static observer, we have $ \left(\partial/\partial r\right)^{b}u^{a}\nabla_{a}u_{b}=\Phi' $.
So we get
\begin{equation}\label{gtt}
\Phi'=-\frac{p'}{\rho+p} \, .
\end{equation}
By substituting this expression into Eq.(\ref{rr}), we obtain
\begin{eqnarray}
\frac{p'}{\rho+p}\Bigg{\{}\sum^{[n/2]}_{i=0}\frac{i\alpha_{i}(n-2)!}{(n-1-2i)!}r^{n-1-2i}e^{-2\Psi}\Big(k-e^{-2\Psi}\Big)^{i-1}\Bigg{\}}
\nonumber\\
=-\kappa^{2}_{n}p~r^{n-2}- \sum^{[n/2]}_{i=0}\frac{\alpha_{i}(n-2)!}{2(n-2i-2)!}r^{n-2-2i}\left( k- e^{-2\Psi} \right)^{i}\, ,
\end{eqnarray}
or
\begin{eqnarray}
\label{TOVlovelock}
\frac{\partial m}{\partial \Psi}\frac{p'}{\rho+p}=-\Big[p~\omega_kr^{n-2}+m'-\frac{\partial m}{\partial \Psi}\Psi'\Big]\, .
\end{eqnarray}
Here $\partial m/{\partial \Psi}$ denotes the derivative of the $m$ respect to $\Psi$ with $r$ fixed as a paremeter.  Eq.(\ref{TOVlovelock}) is generalized TOV equation in Lovelock gravity theory. In following subsection, two examples are given to understand this general expression.

\subsection{Examples}
\subsubsection{Einstein gravity}
If we set $\alpha_0=-2\Lambda=(n-1)(n-2)\lambda$ and $\alpha_1=1$, then we get Einstein gravity theory. In this case, the generalized mass function (\ref{masslovelock}) becomes
\begin{equation}
m=\frac{(n-2)\omega_k}{2\kappa_n^2}r^{n-3}\left[\lambda r^2+ (k-e^{-2\Psi})\right]\, .
\end{equation}
So we have
\begin{equation}
\frac{\partial m}{\partial \Psi}
=-2m+\frac{(n-2)\omega_k}{\kappa_n^2} r^{n-3}(\lambda r^2+k)\, ,
\end{equation}
and
\begin{equation}
m'-\frac{\partial m}{\partial \Psi}\Psi'=\frac{1}{r}\left[(n-3)m+\frac{(n-2)\omega_k}{\kappa_n^2}\lambda r^{n-1}\right].
\end{equation}
Thus, we get the TOV equations in $n$-dimensional Einstein gravity with a cosmological constant:
\begin{equation}
\frac{dp}{dr}= -(\rho+p)\cdot F_E[p,m(r),r]\, ,
\end{equation}
where $F_E[p,m(r),r]$ is an algebraic combination  of $p$, $m(r)$ and $r$, which has the form 
\begin{equation}
F_E[p,m(r),r]=\frac{\big[\kappa_n^2 p+(n-2)\lambda\big]\omega_kr^{n-1}+(n-3)\kappa_n^2m(r)}{r\big[(n-2) (\lambda r^2+k)\omega_k r^{n-3}-2\kappa_n^2 m(r)\big]}\, .
\end{equation}
In four dimension without cosmological constant ($n=4$ and $\lambda=0$), we get the usual TOV equation for the system, i.e.,
\begin{equation}
\frac{dp}{dr}=-(\rho+p)\cdot \Bigg{\{}\frac{m(r)+4\pi p~r^3}{r[r-2m(r)]}\Bigg{\}} \, .
\end{equation}
Actually, in this case, only the case with $k=1$ should be considered. So we have $\omega_k=4\pi$. By choosing $G=1$, we get $\kappa_4^2=8\pi$ and then obtain the result above.
\subsubsection{Gauss-Bonnet gravity}
For Gauss-Bonnet gravity theory, we can also get the explicit form of $\partial m/\partial\Psi$ in terms of $m(r)$. In this case, we  set $\alpha_0=-2\Lambda=(n-1)(n-2)\lambda$, $\alpha_1=1$ and $\alpha_2=\alpha$, and get
\begin{eqnarray}
\label{gaussbonnet}
&&e^{-2\Psi}=k+\frac{r^2}{2(n-3)(n-4)\alpha}(1\pm f)\, ,
\end{eqnarray}
where
\begin{equation}
f=\sqrt{1-4(n-3)(n-4)\lambda+8\alpha\frac{(n-3)(n-4)}{(n-2)}\frac{\kappa_n^2m(r)}{\omega_kr^{n-1}}}\, .
\end{equation}
So the solution has two branches. However, only the negative one in (\ref{gaussbonnet}) can reduce to Einstein gravity theory in large $r$ limit. In the following discussion, only this case will be considered. After some calculation, we get
\begin{equation}
\frac{\partial m}{\partial \Psi}= \frac{(n-2)\omega_k}{\kappa_n^2} r^{n-3} f\Bigg[k- \frac{r^2f}{2(n-3)(n-4)\alpha}\Bigg]\, ,
\end{equation}
and
\begin{eqnarray}
m'-\frac{\partial m}{\partial \Psi}\Psi'= \frac{1}{2\kappa_n^2}\Big[2(n-5)\kappa_n^2 \frac{m(r)}{r}+ (n-1)(n-2)\lambda \omega_k r^{n-2} \Big]\nonumber\\
-\frac{(n-2)\omega_kr^{n-2}}{2(n-3)(n-4)\alpha\kappa_n^2}\Big[(n-3)(n-4)(n-5)\lambda + 1- f\Big]\, .
\end{eqnarray}
So we get the explicit form of $\partial m/\partial \Psi$ and $m'-(\partial m/\partial \Psi)\Psi'$ in terms of the mass function $m(r)$, and then the generalized TOV equation in Gauss-Bonnet gravity theory:
\begin{equation}
\frac{dp}{dr}= -(\rho+p)\cdot F_{GB}[p,m(r),r]\, ,
\end{equation}
with
\begin{eqnarray}
F_{GB}[p,m(r),r]=\frac{r\big[(n-3)(n-4)(n-5)\lambda + (1- f)\big]}{f\big[ 2(n-3)(n-4)\alpha k+r^2(1-f)\big]}+\frac{(n-3)(n-4)\alpha}{n-2}\nonumber\\
\times\frac{2\kappa_n^2 p~\omega_k r^{n-1}+2(n-5)\kappa_n^2m(r) + (n-1)(n-2)\lambda \omega_k r^{n-1}}{\omega_k r^{n-2}f\big[ 2(n-3)(n-4)\alpha k+r^2(1-f)\big]}\, .
\end{eqnarray}
Obviously, $F_{GB}[p,m(r),r]$ is explicitly expressed in terms of $m(r)$, $r$ and $p$. This is similar to the one in Einstein gravity theory.
In principle, for general Lovelock gravity theories with $n\le 10$, we also have this conclusion. For these perfect fluid systems, see~\cite{Dadhich:2010qh,Dadhich:2012cv} for relevant discussions.

\section{Generalized TOV equation from maximum entropy principle}

In previous section, we have get the generalized TOV equation in Lovelock gravity by using gravitational equations. Recently, in \cite{gaosijie}, Gao has used the maximum entropy principle for self-gravitating perfect fluid in Einstein theory to deduce the TOV equation to reveal the implicit relationship between  gravity and thermodynamics. The maximum entropy principle can give part of informations of gravitational equations in Einstein theory, i.e., the TOV equation. In this section, we show that such kind of implicit connection also exists in Lovelock theory. Our work cover the result~\cite{gaosijie} except the case with charges.

\subsection{Local thermodynamic relations of perfect fluid} \label{perfect}

We begin with a brief introduction of local thermodynamic relations of perfect fluid. The perfect fluid system satisfies the familiar first law
\begin{equation}
\label{entropy}
dS=\frac{1}{T}dE+\frac{p}{T}dV-\frac{\mu}{T} dN \,.
\end{equation}
The entropy $S$, energy $E$ and particle number $N$ are all
extensive variables. Their  density variables are denoted by $s$, $\rho$ and $n$ respectively. Applying Eq. (\ref{entropy}) to an unit volume, one can easily find the following density relations
\begin{equation}\label{fl}
ds=\frac{1}{T}d\rho-\frac{\mu}{T}dn \,.
\end{equation}
In the following subsection, we will treat $s$ as the function of two independent variables $(\rho, n)$, i.e., we will use general state equation $s=s(\rho,n)$ without imposing any additional constraint between $\rho$ and $n$.
It is also easily to find the fundamental thermodynamics relation
\begin{equation}\label{euler}
Ts+\mu n=\rho+p
\end{equation}
by the extensive property of the entropy (Euler relation).

\subsection{Maximum entropy principle of perfect fluid} \label{entropy principle}

Now, we apply  the maximum entropy principle to this self-gravitating perfect fluid  in Lovelock gravity. The metric of the spacetime for this system is still given  by Eq.(\ref{1}). However, the situation is quite different from the discussion in previous section: here, only time-time component of gravitational equations Eq.(\ref{GT}) (or Eq.(\ref{tt})) will be used as a constraint to find the TOV equation. Other components of Eq.(\ref{GT}) might be viewed as unknown to us.

We have to emphasize here: Generally, to get the TOV equation, one has to use other components (such as the radial-radial component (\ref{rr})) of the gravitational equation (\ref{GT}). If the maximum entropy principle plus the time-time component constraint equation really implies the TOV equation, we can conclude that the maximum entropy principle does contain part of information of gravitational equations in Lovelock gravity theory,  and then there really exists some implicit relationship between gravity and thermodynamics. This suggests that we may also turn the logic around to explain the phenomena of Lovlock gravity by using thermodynamics.

By using the time-time component of the gravitational equation, i.e., $\emph{G}^{~t}_t=\kappa^{2}_{n}\rho$ as a constraint, the
density $\rho$ can be viewed as a function of the mass $m$ by the relation Eq.(\ref{mrho}), i.e. $$\rho=\frac{m'}{~ \omega_k r^{n-2}}\, ,$$
where $m$ is the Misner-Sharp energy which is given in Eq.(\ref{masslovelock}). On any $t=\mathrm{costant}$ hypersurface, the total entropy of this system is
\begin{equation}
\label{totalentropy}
S=\omega_k\int^{R}_{0}s\Big[\rho\big(m'(r)\big),n(r)\Big]e^{\Psi(m(r))}r^{n-2}dr\, ,
\end{equation}
where $s$ is the entropy density of the perfect fluid and $R$ is the radius of the system. Since $\Psi$ is determined by $m$ and $r$ by an algebraic equation through the definition of the Misner-Sharp energy Eq.(\ref{mass}), we can regard $\Psi$ as a function of $m$ and $r$ algebraically, i.e., $\Psi=\Psi(m,r)$.  However, for simplicity, we have omitted the dependence of $\Psi$ upon $r$ in Eq.(\ref{totalentropy}) above.

The entropy of an isolated system never decreases, and the entropy reaches the maximum when the isolated system is in equilibrium. We hope to find the appropriate $m(r)$ and $n(r)$ to make the entropy maximum, and we consider the entropy is maximum under following conditions: (i). The particle number $N$ is fixed in the system. (ii). The Misner-Sharp mass $m(R)$ is fixed. These two conditions will be used soon.

The total particle number $N$ satisfies
\begin{equation}
N=\omega_k\int^{R}_{0}n(r)e^{\Psi(m(r))}r^{n-2}dr\, ,
\end{equation}
where $n(r)$ is the number density (do not confuse with the dimension $n$). The fixing of the total particle number $N$ is a desired condition for maximum entropy principle, so according to the standard method
of Lagrange multipliers, the equation of variation becomes
\begin{equation}
\delta S+\lambda\delta N=0\, ,
\end{equation}
where $\lambda$ is a Lagrange multiplier. The total mass $m(R)$ is fixed and the mass inside the $r=0$ sphere, i.e., $m(0)$, must be zero, which is also fixed, so their variations have to  satisfy
\begin{equation}
\delta m(0)=\delta m(R)=0 \, .
\end{equation}
Additionally, we suppose the number densities at $r=0$ and $r=R$ are also fixed under the variation, i.e.
\begin{equation}
\delta n(0)=\delta n(R)=0\, .
\end{equation}
By defining the ``total lagrangian"
\begin{equation}
L(m,m',n)=s\big(\rho(m'),n\big)e^{\Psi(m)}r^{n-2}+\lambda n e^{\Psi(m)}r^{n-2}\, ,
\end{equation}
the constrained Euler-Lagrange equations are given by
\begin{equation}\label{2}
\frac{\partial L}{\partial n}=0\, ,
\end{equation}
\begin{equation}\label{3}
\frac{d}{dr}\frac{\partial L}{\partial m{'}}-\frac{\partial
L}{\partial m}=0\, .
\end{equation}
Eq.($\ref{2}$) yields
\begin{equation}
\frac{\partial s}{\partial n}+\lambda=0\, .
\end{equation}
By using the relation ($\ref{fl}$), we have
\begin{equation}\label{constraint}
-\frac{\mu}{T}+\lambda=0\, .
\end{equation}
which shows that $\mu/T$ must be a constant for the self-gravitating
fluid.

Now, let us  consider Eq.(\ref{3}). From the Lagrangian, we have
\begin{equation}
\frac{\partial L}{\partial
m}=\frac{\partial\Psi}{\partial m}(s+\lambda n)e^{\Psi(m)}r^{n-2}\, ,
\end{equation}
where $(\partial \Psi/\partial m)$  is the derivative of $\Psi$ respect to $m$ with the coordinate $r$ fixed as a parameter. Further, we have
\begin{equation}
\frac{\partial L}{\partial m'}=\frac{\partial s}{\partial
\rho}\frac{\partial \rho}{\partial
m'}e^{\Psi(m)}r^{n-2}=\frac{e^{\Psi(m)}}{T\omega_k},
\end{equation}
(where Eq.(\ref{fl}) and Eq.(\ref{mrho}) have been used) and
\begin{equation}
\frac{d}{dr}\frac{\partial L}{\partial
m'}=\Big(\Psi'-\frac{T'}{T}\Big)\frac{e^{\Psi(m)}}{T\omega_k} \,.
\end{equation}
By using the EOM of $n$ and the fundamental relation  ($\ref{euler}$), one can write the
Euler-Lagrangian equation of $m$ in an explicit form
\begin{equation}\label{eomofm}
\frac{T{'}}{T}=-\omega_k(\rho+p)r^{n-2}\frac{\partial \Psi}{\partial m}+\Psi'\, .
\end{equation}
The constraint ($\ref{constraint}$) yields
\begin{equation}
\mu'=\lambda T'.
\end{equation}
Considering the fundamental relation of thermodynamics ($\ref{euler}$) and Eq.($\ref{fl}$), we have
\begin{equation}
dp=sdT+nd\mu.
\end{equation}
It follows immediately that
\begin{equation}
p'(r)=sT'+n\mu'=T'(s+\lambda n)=\frac{T'}{T}(\rho+p)\, .
\end{equation}
Thus we obtain an equation
\begin{equation}
\label{TOV}
\frac{p'}{\rho+p}=-\omega_k(\rho+p)r^{n-2}\frac{\partial \Psi}{\partial m}+\Psi'.
\end{equation}
Considering $(\partial \Psi/ \partial m)(\partial m/\partial\Psi)=1$, we have
\begin{equation}
\label{TOVLovelock1}
\frac{\partial m}{\partial \Psi}\frac{p'}{\rho+p}=-\Big[\omega_k(\rho+p)r^{n-2}-\frac{\partial m}{\partial \Psi}\Psi'\Big]\, .
\end{equation}
Remembering $m'=\rho~\omega_k r^{n-2} $, this equation is nothing but the generalized TOV equation (\ref{TOVlovelock}) in Lovelock gravity theory.

Here we have successfully deduced the TOV equation only using the maximum entropy principle and the time-time component gravitational equation as a constraint. Now, we can say that the maximum entropy principle really gives part of information of gravitational equations and the implicit connection between gravity and thermodynamics also exists  in Lovelock gravity. For the cases with charges, this discussion can be performed straightforwardly, and we will not discuss them here.

\section{Conclusions and Discussion}

The TOV equation is an important equation for self-gravitating system which was originally derived from the Einstein equation. We have got the generalized TOV equation in Lovelock gravity theory. This equation has a general form $$\frac{dp}{dr}=-(\rho+p)\cdot F[p,m(r),r]$$ with a different factor $F[p,m(r),r]$ for a different gravity theory. By applying the maximum entropy principle to a general self-gravitating fluid, we have also derived the TOV equation of hydrostatic equilibrium  only using the time-time component of the gravitational constraint equation and ordinary thermodynamic relations. Our results show that  part of the gravitational equations of Lovelock theory  can also be derived from ordinary thermodynamic laws. This is a direct evidence for the possible fundamental relationship between gravitation and thermodynamics.

It might be interesting to discuss the problem in a more ambitious way: Perhaps,
one may think that the time-time component of the gravitational equations is  not necessary!
For the same system described at the beginning of Sec.3, we can define a mass function by $$m(r)=\omega_k\int_0^R \rho r^{n-2}dr\, , $$
then we have $m'= \rho \omega_k r^{n-2}$. Further, $\Psi$ is assumed as a function of $m$. Then, by these assumptions and the maximum entropy principle, following the same reduction in Sec.4, we get Eq.(\ref{TOVLovelock1}) or Eq.(\ref{TOVlovelock}). Thus, the remainder unknown information is just the relation between $m$ and $\Psi$. This is determined by the structures of various gravity theories. So even without the time-time component of gravitational equations, we can also get some information about the TOV equation: At least, it should has a form like Eq.(\ref{TOVlovelock}). However, without the details of the time-time component, we can not fixed the ambiguity in Eq.(\ref{TOVlovelock}) (i.e., the detailed relation between $m$ and $\Psi$). So it seems that time-time component of the gravitational equations is indispensable to get the TOV equation.

It is interesting to apply this principle to other generalized gravity theories and investigate whether this deep connection between gravity and thermodynamics exits or not. Another important question is: whether it is possible to apply the maximum entropy principle to some dynamical system (time dependent system). The situation becomes difficult once a dynamical system is considered. Actually, we do not know how to define an equilibrium state in this case. It might be necessary to improve the maximum entropy principle to solve such problem, which needs further discussion.

\section{Acknowledgement}
This work is supported by NSFC Grants No.11205148 and No.11235010. LMC would like to thank Rong-Gen Cai for his useful discussion and kindly help. We also would like to thank Sijie Gao for his useful discussions and comments.

\end{CJK*}  
\end{document}